\documentclass[conference]{IEEEtran}
\IEEEoverridecommandlockouts
\usepackage{cite}
\usepackage{amsmath,amssymb,amsfonts}
\usepackage{algorithm,algorithmic}
\usepackage{graphicx}
\usepackage{textcomp}
\usepackage{xcolor}
\usepackage{bm,bbm}
\usepackage{color}
\def\BibTeX{{\rm B\kern-.05em{\sc i\kern-.025em b}\kern-.08em
    T\kern-.1667em\lower.7ex\hbox{E}\kern-.125emX}}

\newcommand{\ip}[2]{\left\langle#1,#2\right\rangle}
\newcommand{\norm}[1]{\left\lVert#1\right\rVert}

\renewcommand{\[}{\left[}
\renewcommand{\]}{\right]}

\newcommand{\ve}{\bm{e}}
\newcommand{\vg}{\bm{g}}

\newcommand{\vn}{\bm{n}}

\newcommand{\vx}{\bm{x}}
\newcommand{\vu}{\bm{u}}
\newcommand{\vv}{\bm{v}}
\newcommand{\vw}{\bm{w}}
\newcommand{\vy}{\bm{y}}
\newcommand{\vz}{\bm{z}}

\newcommand{\vA}{\bm{A}}

\newcommand{\vI}{\bm{I}}

\newcommand{\R}{\mathbbm{R}}
\newcommand{\bS}{\mathbb{S}}
\newcommand{\bT}{\mathbb{T}}
\newcommand{\bL}{\mathbb{L}}

\newtheorem{theorem}{Theorem}

\newtheorem{fact}{Fact}

\newtheorem{lemma}{Lemma}

\newenvironment{proofsketch}{{\noindent\it \textbf{Sketch of Proof}.}\quad}{\hfill $\square$\par}

\begin{document}

\title{Data-Time Tradeoffs for Optimal $k$-Thresholding Algorithms in Compressed Sensing\\
\thanks{This work was supported by Postdoctoral International Exchange Program for Incoming Postdoctoral Students under Grant No. YG20200197 and China National Postdoctoral Program for Innovative Talents under Grant No. BX2021346. (Correpsonding author: Xu Zhang.) A full version of this paper is accessible at:
\texttt{https://arxiv.org/abs/2110.06460}.}
}

\author{\IEEEauthorblockN{Jialiang Xu}
\IEEEauthorblockA{\textit{LSEC, Academy of Mathematics and Systems Science} \\
\textit{Chinese Academy of Sciences}\\
Beijing, China \\
xujialiang@lsec.cc.ac.cn}
\and
\IEEEauthorblockN{Xu Zhang}
\IEEEauthorblockA{\textit{LSEC, Academy of Mathematics and Systems Science} \\
\textit{Chinese Academy of Sciences}\\
Beijing, China \\
xuzhang\_cas@lsec.cc.ac.cn}
}

\maketitle

\begin{abstract}
Optimal $k$-thresholding algorithms are a class of $k$-sparse signal recovery algorithms that overcome the shortcomings of traditional hard thresholding algorithms caused by the oscillation of the residual function. 
In this paper, a novel convergence analysis for optimal $k$-thresholding algorithms is established, which reveals the data-time tradeoffs of these algorithms.
Both the analysis and numerical results demonstrate that when the number of measurements is small, the algorithms cannot converge; when the number of measurements is suitably large,  the number of iterations required for successful recovery has a negative correlation with the number of measurements, and the algorithms can achieve linear convergence.  Furthermore, the main theorems indicate that the number of measurements required for successful recovery is of the order of $k \log({n}/{k})$, where $n$ is the dimension of the target signal.
\end{abstract}


\section{Introduction}
\label{sec:intro}

Compressed sensing (CS) together with its sparse optimization models has attracted plenty of attention in signal and image processing over the past few decades \cite{candes2005,donoho2006,candes2006}.
CS concerns the problem of recovering a sparse signal from a relatively small number of noisy measurements \begin{equation}
    \vy=\vA \vx^\star+\vn,
\end{equation}
where $\vx^\star\in \mathbbm{R}^n$ is the target signal, $\vA\in \mathbbm{R}^{m\times n}\, (m \le n)$ is the sensing matrix, and $\vn\in \mathbbm{R}^m$ is the noise. Finding the target signal from the above underdetermined system can be formulated as the following optimization problem with a sparsity constraint
\begin{equation}\label{model:p1}
    \min_{\vx}\left\{\|\vy-\vA\vx\|_2^2:  \norm{ \vx}_0\le k\right\},
\end{equation}
where $\norm{\vx}_0$ denotes the number of nonzero entries of the vector $\vx$, $k \ll n $ is an integer number. Although the problem \eqref{model:p1} is NP-hard \cite{bk1995}, it can be solved by some efficient algorithms with high probability. 

The plausible algorithms to solve this problem have been classified into three categories:  convex optimization \cite{chen2001,candes2008,zhao2012}, heuristic methods \cite{mp,sp}, and thresholding methods \cite{softth,hardth,zhao2020optimal}. Here we focus on the thresholding methods. Generally speaking, thresholding methods can be classified into soft thresholding \cite{softth}, hard thresholding \cite{hardth}, and optimal $k$-thresholding \cite{zhao2020optimal}.  The soft thresholding algorithms are established by considering the optimality condition of certain convex problems. To describe hard thresholding methods, it is convenient to introduce the hard thresholding operator $\mathcal{H}_k(\vz)$, which retains the $k$ largest absolute entries of $\vz$ and zeroes out the remaining entries. 
By connecting $\mathcal{H}_k(\vz)$ and the gradient method for minimizing the  residual  function $\| \vy-\vA \vx\|_2^2$, Blumensath and Davies \cite{hardth} proposed the so-called iterative hard thresholding (IHT) to solve the problem \eqref{model:p1}, which takes the scheme
\begin{equation}\label{IHT}
    \vx^{p+1}=\mathcal{H}_{k}(\vx^p+\vA^{T}(\vy-\vA \vx^{p})).
\end{equation}
 By combining IHT and  the pursuit step such as minimizing the residual subject to the support determined by IHT, Foucart \cite{HTP} put forward an accelerated version called hard thresholding pursuit (HTP). However, in many numerical situations, it demonstrates that the residual reduction is irrelevant \textcolor{black}{to} the thresholding step, causing the divergence of IHT, \textcolor{black}{and} even HTP. It is worth mentioning that Zhao \cite{zhao2020optimal} proposed an iterative scheme to seek optimal $k$ locations of $\vu^p$ that minimize the residual reduction before the thresholding step
\begin{equation}\label{scheme}
    \vw(\vu^p)= \mathrm{min}   \big{\{}\|\vy- \vA (\vu^p  \odot   \vw) \|^2_2, ~ \textrm{\textbf{e}}^T \vw= k , ~    \vw\in \left\{ 0,1\right\}^n\big{\}},
\end{equation}
and  update the next iteration with $\vx^{p+1}=\vu^p \odot \vw(\vu^p)$, where $\vu^p=\vx^p+\vA^{T}(\vy-\vA \vx^{p})$, $\textrm{\textbf{e}} \in \R^n$ denotes all-one vector and the symbol $\odot$ denotes Hadamard product. This algorithm is called optimal $k$-thresholding (OT) algorithm and its variant with pursuit step over the support of $ \vw(\vu^p)$ is called optimal $k$-thresholding pursuit (OTP) algorithm. To make the algorithms tractable, Zhao studied their convex relaxation versions by replacing $\vw\in \left\{ 0,1\right\}^n$ with $0\leq \vw\leq \textrm{\textbf{e}}$, i.e., ROT and ROTP algorithms (see Algorithm \ref{alg: I} and Algorithm \ref{alg: II} for more details). Numerical experiments have shown that ROP or ROTP algorithms are more robust and stable during the course of iterations, and have better performance on signal recovery than traditional hard thresholding methods \cite{zhao2020optimal}.   

Under the assumption that the sensing matrix satisfies the restricted isometry property (RIP), Zhao \cite{zhao2020optimal} has shown that the restricted isometry constant $\delta_{3k}<1/5$ is sufficient for the convergence of both ROT and ROTP. However, the main drawback of using RIP in convergence analysis for the algorithms is that
RIP-based convergence analysis cannot characterize the relationship among reconstruction accuracy, the number of measurements and the number of iterations, which is of great importance in the field related with data acquisition from underdetermined systems \cite{oymak2017,chen2018}. In particular, the theoretical analysis of algorithms is required to consider and understand the precise tradeoffs between the number of obtainable measurements and the convergence speed, which we called the data-time tradeoffs for algorithms. The researches in \cite{cj2013,btcb2015} showed that more efficient algorithms can be obtained by the use of hierarchies of convex relaxation when the number of measured data is relatively large. The sharp data-time tradeoffs for projected gradient methods to solve linear inverse problems were analyzed by Oyamk, Recht, and Soltanolkotabi \cite{oymak2017}. Chen and Liu \cite{chen2018} characterized the data-time tradeoffs for projected gradient descent algorithms in corrupted  sensing under sub-Gaussian measurements.


To overcome the drawback, this paper establishes a novel  convergence analysis for the relaxed optimal $k$-thresholding algorithms ROT and ROTP, and provides the data-time tradeoffs for these algorithms under sub-Gaussian measurements. Theoretical analysis presents that when the number of measurements is relatively large, with the increase of the number of measurements, the number of iterations needed for successful reconstruction decreases accordingly. Moreover, if the number of measurements is of the order of $k \log({n}/{k})$ with a relatively large constant, we show that the algorithms can achieve linear convergence rate.

The paper is organized as follows. Section  \ref{sec:format} serves as an introduction to some necessary definitions and facts. Section \ref{sec:TheoreticalGuarantees} presents the theoretical analysis (considering the data-time tradeoffs) for ROT and ROTP algorithms. The numerical results are provided in Section \ref{sec:Simulations}. Finally, the conclusion and future work are given in Section \ref{sec:Conclusion}.

\begin{algorithm}[!t]
	\caption{Relaxed optimal $k$-thresholding (ROT) algorithm}
	\label{alg: I}
	\begin{algorithmic}[1]
		\REQUIRE  Matrix $\vA$, measurements $\vy$,  sparsity level $k$, maximum iteration number $P$
		\ENSURE   Estimated result $\vx^{\star}$
		\STATE
		Initialize
		$p=0,$ $\varepsilon,$ $\eta,$ and $\vx^{0}$
		\REPEAT
		\STATE {$\vu^p=\vx^p+\eta \vA^T(\vy-\vA \vx^p)$}
		\STATE $\vw(\vu^p)= \mathrm{argmin}_{\vw}    \big{\{}\|\vy- \vA (\vu^p  \odot   \vw) \|^2_2, ~ \textrm{\textbf{e}}^T \vw= k , ~   0\leq \vw\leq \textrm{\textbf{e}}\big{\}}$
		\STATE $\vx^{p+1} =  {\cal H}_k (\vu^p  \odot  \vw(\vu^p))$
		\STATE $p=p+1$
		\UNTIL $p>P$ or $\norm{{\vx^p-\vx^{p-1}}}_2<\varepsilon$
	\end{algorithmic}
\end{algorithm}

\section{Preliminaries}
\label{sec:format}
In this section, we give several definitions and facts, which are useful in the following sections. All the definitions can be found in \cite{vershynin2018high}.



\subsection{Useful definitions}

A random variable $x$ is called \emph{sub-Gaussian} if it has a finite sub-Gaussian norm 
\begin{equation} \label{eq: subGaussianvariable}
    \norm{x}_{\psi_2}=\inf\left\{t>0: \textbf{E} \left[\exp\left(\frac{x^2}{t^2}\right)\right]\le 2\right\}.
\end{equation}
For example, Gaussian, Bernoulli, and bounded random variables are typical sub-Gaussian variables. A random vector $\vx \in \mathbbm{R}^n$ is called \emph{sub-Gaussian}  if for any $\vy \in \mathbbm{R}^n$, $\ip{\vx}{\vy}$ is sub-Gaussian. The corresponding norm is defined as 
\begin{equation} \label{eq: subGaussianvector}
    \norm{\vx}_{\psi_2}=\sup_{\vy \in \mathbbm{S}^{n-1}} \norm{\ip{\vx}{\vy}}_{\psi_2},
\end{equation}
where $\mathbbm{S}^{n-1}$ denotes the $n$-dimensional unit sphere.

A random vector $\vx$ is \emph{isotropic} if it satisfies 
\begin{equation} \label{eq: isotropic}
\textbf{E} [\vx \vx^T]=\bm{I}_n,
\end{equation}
where $\bm{I}_n$ denotes the $n$-dimensional identity matrix.

The \emph{Gaussian complexity} of a bounded subset $\mathbb{T}$ is defined as 
  $ \gamma(\mathbb{T})=\textbf{E} \Big{[}\sup_{\vx \in \mathbb{T}} |\ip{\vg}{\vx}|\Big{]},$
and the \emph{Gaussian width} of a bounded subset $\mathbb{T}$ is defined as 
$   w(\mathbb{T})=\textbf{E}\Big{[}\sup_{\vx \in \mathbb{T}} \ip{\vg}{\vx}\Big{]},$
where $\vg \sim \mathcal{N}(\bm{0},\vI_n)$.
The above two quantities are of the same order of magnitude \cite{chen2018stable} due to the fact
\begin{equation}\label{eq:cworder}
\frac{1}{3}(w(\mathbb{T})+\|\boldsymbol{y}\|_{2}) \leq \gamma(\mathbb{T}) \leq 2\left(w(\mathbb{T})+\|\boldsymbol{y}\|_{2}\right),  \forall \, \boldsymbol{y} \in \mathbb{T}.
\end{equation}

\begin{algorithm}[!t]
	\caption{Relaxed optimal $k$-thresholding pursuit (ROTP) algorithm}
	\label{alg: II}
	\begin{algorithmic}[1]
		\REQUIRE  Matrix $\vA$, measurements $\vy$,  sparsity level $k$, maximum iteration number $P$
		\ENSURE   Estimated result $\vx^{\star}$
		\STATE
		Initialize
		$p=0,$ $\varepsilon,$ $\eta,$ and $\vx^{0}$
		\REPEAT
		\STATE {$\vu^p=\vx^p+\eta \vA^T(\vy-\vA \vx^p)$}
		\STATE $\vw(\vu^p)= \mathrm{argmin}_{\vw}    \big{\{}\|\vy- \vA (\vu^p  \odot   \vw) \|^2_2,~ \textrm{\textbf{e}}^T \vw= k , ~   0\leq \vw\leq \textrm{\textbf{e}}\big{\}}$
		\STATE $\vv =  {\cal H}_k (\vu^p  \odot  \vw(\vu^p))$
		\STATE $\mathbb{S}^{p+1}=\mbox{supp}(\vv)$
		\STATE $\vx^{p+1} = \mathrm{argmin}_{\vx} \big{\{}\|\vy- \vA \vx \|^2_2, ~ \mbox{supp}(\vx) \subseteq \mathbb{S}^{p+1} \big{\}}$
		\STATE $p=p+1$
		\UNTIL $p>P$ or $\norm{{\vx^p-\vx^{p-1}}}_2<\varepsilon$
	\end{algorithmic}
\end{algorithm}

\subsection{Useful facts}
\begin{fact}[Matrix deviation inequality, \cite{vershynin2017four}] \label{lm:MatrixDeviationInequality}
 Let $\vA$ be an ${m \times n}$ random matrix whose rows $\{\vA_{i}\}_{i=1}^m$ are independent, centered, isotropic and sub-Gaussian vectors. For any bounded subset $\mathbb{D} \subseteq \mathbbm{R}^n$ and $t \ge 0$, the inequality
$$
    \sup \limits_{\vx \in \mathbb{D}} \left| \norm{\vA \vx}_2 - \sqrt{m} \norm{\vx}_2 \right| \le CK^2 [\gamma (\mathbb{D})+t\cdot \mbox{rad}(\mathbb{D})]
$$
holds with probability at least $1-2\exp(-t^2)$. Here, $ \mbox{rad}(\mathbb{D})=\sup_{\vx \in \mathbb{D}} \norm{\vx}_2$, $K=\max_i \norm{\vA_{i}}_{\psi_2}$ and $C>0$ is an absolute constant.
\end{fact}
\begin{fact}[Gaussian width of $\ell_0$-balls, \cite{wainwright2019high}]\label{fact2} Let $\mathbb{D}_{k}= \{\vz \in \mathbbm{R}^n \big{|} \norm{\vz}_0 \le k, \norm{\vz}_2 = 1\}$. Then its Gaussian width satisfies
\begin{align}
    w(\mathbb{D}_{k}) \le c \sqrt{k \log\frac{en}{k}}.
\end{align}
\end{fact}

\section{Theoretical Guarantees}
\label{sec:TheoreticalGuarantees}
For the convenience of the convergence analysis, we first give some useful lemmas, which will be frequently used in our later analysis (Theorems \ref{Thm_ROT} and \ref{Thm_ROTP}). Due to the space limit, we only give the sketch of the proofs, and the complete proofs are presented in the full version. 

By assigning $\mathbb{D}\in \mathbbm{S}^{n-1}$ and $t=\gamma(\mathbb{D})$ from Fact \ref{lm:MatrixDeviationInequality}, we have the following lemma.

\begin{lemma} Let $\vA$ be an ${m \times n}$ random matrix as defined in Fact \ref{lm:MatrixDeviationInequality}. Let $\mathbb{D}\in \mathbbm{S}^{n-1}$ and $t=\gamma(\mathbb{D})$, then for any $\vx \in \mathbb{D}$, the inequality 
    $$
     \sqrt{m} - 2CK^2 \gamma (\mathbb{D}) \le \norm{\vA \vx}_2  \le \sqrt{m}+ 2CK^2 \gamma (\mathbb{D})
    $$
    holds with probability at least $1-2\exp(-\gamma^2(\mathbb{D}))$.
\end{lemma}
\begin{lemma}\label{lm:OTROT}
    Let $\vA$ be an ${m \times n}$ random matrix as defined in Fact \ref{lm:MatrixDeviationInequality}.  Let $\mathbb{D}_k= \{\vz  \in \R^n \big{|} \|\vz\|_0 \le k, \|\vz\|_2 = 1\}$. Let $\mathbb{L} \subseteq \{1,\ldots,n\}$ be an index set with cardinality $l$. Define $\left[ \vx\right]_{\mathbb{L}}$ as the vector obtained by retaining the components of $\vx$ with the support $\mathbb{L}$ and zeroing out the remaining components. Let $\eta$ be a step size such that $\eta  m \le 1$. Then we have 
     
    \noindent\textbf{a)}  For any $k$-sparse vector $\vv  \in \mathbbm{R}^n$, with  probability at least $1-4\exp(-\gamma^2(\mathbb{D}_k+\mathbb{D}_l))$, we have $\norm{[(\vI-\eta \vA^T \vA)\vv]_\mathbb{L}}_2 \le \max\left\{s_1, s_2 \right\} \cdot \norm{\vv}_2$, where
    \begin{align*}
      s_1&=1-\eta m \cdot    \frac{\sqrt{m}-6CK^2\gamma(\mathbb{D}_k+\mathbb{D}_l)}{\sqrt{m}},\\
     s_2&=\frac{3CK^2\gamma(\mathbb{D}_k+\mathbb{D}_l)}{\sqrt{m}}\left(\frac{3CK^2\gamma(\mathbb{D}_k+\mathbb{D}_l)}{2\sqrt{m}}+1\right).  
    \end{align*}
        
    \noindent\textbf{b)}  For any $\vx\in \mathbbm{R}^m$, with probability at least $1-2\exp(-\gamma^2(\mathbb{D}_l))$, we have $ \norm{[\vA^T \vx]_\mathbb{L}}_2 \le  \norm{\vx}_2(\sqrt{m} +2CK^2\gamma(\mathbb{D}_l))$.
\end{lemma}
\begin{proofsketch} \noindent\textbf{a)} Reformulate $ \norm{[(\vI - \eta  \vA^T \vA) \vv]_\mathbb{L} }_2$ as 
\begin{align}
    \norm{[(\vI - \eta  \vA^T \vA) \vv]_\mathbb{L} }_2&=
    \ip{[(\vI-\eta \vA^T \vA)\vv]_\mathbb{L} }{\bm{\mu}} \nonumber\\
    &=\norm{\vv}_2\ip{(\vI - \eta \vA^T \vA)  \bm{\nu} }{\bm{\mu}},
\end{align}
where
\begin{equation*}
    \bm{\mu}=\frac{[(\vI-\eta \vA^T \vA)\vv]_\mathbb{L} }{\norm{[(\vI - \eta  \vA^T \vA) \vv]_\mathbb{L} }_2} \in \mathbb{D}_l,~
    \bm{\nu}=\frac{\vv}{\norm{\vv}_2} \in \mathbb{D}_k.
\end{equation*}
As for $\ip{(\vI - \eta  \vA^T \vA) \bm{\nu}}{\bm{\mu}}$, we have 
\begin{align}
    \ip{(\vI - \eta  \vA^T \vA) \bm{\nu}}{\bm{\mu}} =&\frac{1}{4}\Bigg{(}\norm{ \bm{\nu}+\bm{\mu}}_2^2 -\eta  \norm{\vA( \bm{\nu}+\bm{\mu})}_2^2
    \nonumber\\
    & -\norm{ \bm{\nu}-\bm{\mu}}_2^2+\eta \norm{\vA( \bm{\nu}-\bm{\mu})}_2^2\Bigg{)}.
\end{align}
Using Lemma \ref{lm:MatrixDeviationInequality} yields the bounds 
\begin{multline}
    \norm{\vA( \bm{\nu}+\bm{\mu})}_2^2 \\ \ge \left(\max\{\sqrt{m} \norm{ \bm{\nu}+\bm{\mu}}_2 -3CK^2\gamma(\mathbb{D}_k+\mathbb{D}_l),0\}\right)^2
\end{multline}
and
\begin{equation}
    \norm{\vA( \bm{\nu}-\bm{\mu})}_2^2 \le \left[\sqrt{m}\norm{ \bm{\nu}-\bm{\mu}}_2 +3CK^2\gamma(\mathbb{D}_k+\mathbb{D}_l)\right]^2
\end{equation}
with probability at least $1-2\exp(-\gamma^2(\mathbb{D}_k+\mathbb{D}_l))$ respectively.

Then we give the upper bound of $\ip{(\vI - \eta  \vA^T \vA) \bm{\nu}}{\bm{\mu}}$ in two cases: $\sqrt{m} \norm{ \bm{\nu}+\bm{\mu}}_2-3CK^2\gamma(\mathbb{D}_k+\mathbb{D}_l)\ge 0$ and $\sqrt{m} \norm{ \bm{\nu}+\bm{\mu}}_2 -3CK^2\gamma(\mathbb{D}_k+\mathbb{D}_l)< 0$. After that, we complete the proof.

\textbf{b)} Reformulate $\norm{[\vA^T \vx]_\mathbb{L}}_2$ as
\begin{align} 
    \norm{[\vA^T \vx]_\mathbb{L}}_2&=\ip{[\vA^T \vx]_\mathbb{L} }{\vz} 
    =\ip{\vA^T \vx }{\vz} 
    =\ip{\vx}{\vA \vz},
\end{align}
where
\begin{equation*}
    \vz=\frac{[\vA^T \vx]_\mathbb{L} }{\norm{[\vA^T \vx]_\mathbb{L}}_2} \in \mathbb{D}_l.
\end{equation*}
Using H\"{o}lder's inequality and  Lemma \ref{lm:MatrixDeviationInequality} completes the proof
\begin{equation*}
    \norm{[\vA^T \vx]_\mathbb{L}}_2\le \norm{\vx}_2 \norm{\vA \vz}_2 \nonumber\le \norm{\vx}_2(\sqrt{m} +2CK^2\gamma(\mathbb{D}_l)).
\end{equation*} 
\end{proofsketch}

\begin{lemma} \label{lm:Dk} Let $\mathbb{D}_{k}= \{\vz \in \R^n \big{|} \|\vz\|_0 \le k, \|\vz\|_2 = 1\}$. Then we have

\noindent\textbf{a)}  $\label{gamma_ineq_1}
    \gamma(\mathbb{D}_{2k}) \le 2 \gamma(\mathbb{D}_{k}),
$

\noindent\textbf{b)} $  \label{gamma_ineq_2}
    \gamma(\mathbb{D}_{k}+\mathbb{D}_{k}) \le 2 \gamma(\mathbb{D}_{k}),
$

\noindent\textbf{c)} $ \label{gamma_ineq_3}
    \mathbb{D}_{k} \subseteq \mathbb{D}_{l}$ and $\gamma(\mathbb{D}_{k}) \le \gamma(\mathbb{D}_{l})$ for any $k\le l$.

\end{lemma}

Now we show the convergence analysis for the ROT algorithm based on the above lemmas.
\begin{theorem}\label{Thm_ROT}
Let $\vA$ be an ${m \times n}$ random matrix as defined in Fact \ref{lm:MatrixDeviationInequality}. Let $\vx^\star\in \mathbbm{R}^n$ be a $k$-sparse vector such that $\vy=\vA \vx^\star+\vn$.  Let $\eta$ be the step size satisfying $\eta m \le 1$.
Then for any sequence $\left\{ \vx^p\right\}$ generated by ROT, one has 
\begin{equation}\label{Re3}
    \norm{ \vx^\star-\vx^{p}}_2\le \rho_1 \norm{ \vx^\star-\vx^{p-1}}_2+ c_2 \norm{\vn}_2
\end{equation}
with probability $1-c\exp(-\gamma^2(\mathbb{D}_{2k}))$, where $c>0$ is a constant and
\begin{align*}
    \rho_1=& c_1 \max\left\{r_1,r_2 \right\},~~ c_2=c_{21}+c_{22},\\
    r_1=&1-\eta m \cdot    \frac{\sqrt{m}-12CK^2\gamma(\mathbb{D}_{2k})}{\sqrt{m}},\\ r_2=&\frac{6CK^2\gamma(\mathbb{D}_{2k})}{\sqrt{m}}\left(\frac{3CK^2\gamma(\mathbb{D}_{2k})}{\sqrt{m}}+1\right),\\
    c_1= &\frac{4\sqrt{m}+4CK^2\gamma(\mathbb{D}_{k})}{\sqrt{m}-2CK^2\gamma(\mathbb{D}_{2k})}, \\
    c_{21}=&\frac{3\eta (\sqrt{m}+2CK^2\gamma(\mathbb{D}_{k}))^2+2}{\sqrt{m}-2CK^2\gamma(\mathbb{D}_{2k})},\\
    c_{22}=& \eta(\sqrt{m}+2CK^2\gamma(\mathbb{D}_{k})).
\end{align*}
Let $m$ be suitably large satisfying $\rho_1< 1$. Then  $\left\{ \vx^p\right\}$ can approximate $\vx^\star$ with the error
\begin{equation}\label{Re4}
    \| \vx^\star-\vx^{p}\|_2\le \rho_1^p \| \vx^\star-\vx^{0}\|_2+ \frac{c_2}{1-\rho_1}  \norm{\vn}_2
\end{equation}
with probability $1-c'\exp(-\gamma^2(\mathbb{D}_{2k}))$,  where $c'>0$ is a constant. In particular, when $\vn=\bm{0}$, the sequence $\left\{ \vx^p\right\}$ converges to $\vx^\star$.
\end{theorem}

\begin{proofsketch}
The main step is to bound   $\| \vx^\star-\vx^{p}\|_2$ by using the inequality in \cite{zhao2020optimal}: 
\begin{align}
 \|\vx^{\star}-\vx^{p}\|_2&\leq \norm{ \left[ \vu^{p-1} \odot \vw(\vu^{p-1})-\vx^{\star} \right]_{\mathbb{S}^{p}\setminus \mathbb{S}}}_2, \nonumber\\
&\qquad +\norm{\left[ \vu^{p-1} \odot \vw(\vu^{p-1})-\vx^{\star} \right]_{\mathbb{S}^{p}\cup \mathbb{S}}}_2,
\end{align}
where   $\bS^{p}$ and $\bS$ are the support of $\vx^{p}$ and $\vx^\star$, respectively.

First, we give the upper bound of $\| [ \vu^{p-1} \odot \vw(\vu^{p-1})-\vx^{\star} ]_{\mathbb{S}^{p}\setminus \mathbb{S}}\|_2$. By using $\vu^{p-1}=\vx^{p-1}+\eta \vA^T(\vy-\vA \vx^{p-1})$ and $\vy=\vA\vx^\star+\vn$, we obtain
\begin{equation}\label{transaction_a}
    \vx^\star-\vu^{p-1}  =(\vI - \eta  \vA^T \vA) (\vx^\star-\vx^{p-1})-\eta  \vA^T \vn.
\end{equation}
Due to the fact that $\vx^\star_{\bS^{p}\setminus \bS}=0$, we have
\begin{align}
     &\norm{ \left[\vu^{p-1} \odot \vw(\vu^{p-1})-\vx^{\star}\right]_{\mathbb{S}^{p}\setminus \mathbb{S}}}_2 \nonumber\\
    =&\norm{ \left[(\vu^{p-1}-\vx^\star) \odot \vw(\vu^{p-1})\right]_{\bS^{p}\setminus \bS}}_2.  
\end{align}  
Using \eqref{transaction_a}, the triangle inequality, and $(\vw(\vu^{p-1}))_i\in [0,1],i=1,\ldots,n$ yields that
\begin{multline}  
   \norm{ \left[\vu^{p-1} \odot \vw(\vu^{p-1})-\vx^{\star}\right]_{\mathbb{S}^{p}\setminus \mathbb{S}}}_2
    \le \eta \norm{ \left[  \vA^T \vn  \right]_{\bS^{p}\setminus \bS}}_2 
    \\+ \norm{ \left[(\vI - \eta  \vA^T \vA) (\vx^\star-\vx^{p-1}) \right]_{\bS^{p}\setminus \bS}}_2,\label{eq2:ROT}
\end{multline}
By using Lemmas \ref{lm:OTROT} and \ref{lm:Dk}, we get 
\begin{multline}\label{eq5:ROT}
    \norm{\left[\vu^{p-1} \odot \vw(\vu^{p-1})-\vx^\star\right]_{\bS^{p}\setminus \bS}}_2\le \Big{[}\max\left\{r_1,r_2 \right\}  \\  \cdot \norm{\vx^\star-\vx^{p-1}}_2 \Big{]} 
    +\eta \norm{\vn}_2(\sqrt{m} +2CK^2\gamma(\mathbb{D}_k)).
\end{multline}

Next, we provide the upper bound of $\|[ \vu^{p-1} \odot \vw(\vu^{p-1})-\vx^{\star} ]_{\mathbb{S}^{p}\cup \mathbb{S}}\|_2$. Note that 
\begin{align}  \label{eq7:ROT}
    & \vy-\vA\left[\vu^{p-1} \odot \vw(\vu^{p-1})\right] \nonumber\\
    =  & \vA\left[\vu^{p-1} \odot \vw(\vu^{p-1})-\vx^\star\right]-\vn \nonumber\\ 
    =  & \vA\left[\vu^{p-1} \odot \vw(\vu^{p-1})-\vx^\star\right]_{\bS^{p}\cup \bS}\quad \nonumber\\
    & + \vA\left[\vu^{p-1} \odot \vw(\vu^{p-1})-\vx^\star\right]_{\overline{\bS^{p}\cup \bS}}-\vn
\end{align}
where the first equality uses $\vy=\vA\vx^\star+\vn$. By using the triangle inequality and Lemma \ref{lm:MatrixDeviationInequality}, we have 
\begin{multline}\label{eq8:ROT}
    \norm{ \left[\vu^{p-1} \odot \vw(\vu^{p-1})-\vx^\star\right]_{\bS^{p}\cup \bS}}_2 \\\le \frac{1}{\sqrt{m}-2CK^2\gamma(\mathbb{D}_{2k})} 
    \cdot \left\{ U+T+\|\vn\|_2 \right\},
\end{multline}
where $U=\norm{ y-\vA\left[\vu^{p-1} \odot \vw(\vu^{p-1})\right]}_2$ and $T=\norm{ \vA\left[\vu^{p-1} \odot \vw(\vu^{p-1})-\vx^\star\right]_{\overline{\bS^{p}\cup \bS}}}_2$. It follows from the optimality of $\vw(\vu^{p-1})$ to the residual function, Lemma \ref{lm:MatrixDeviationInequality}, and Lemma \ref{lm:OTROT} that
 \begin{multline}\label{eq3}
     U\le (\sqrt{m}+2CK^2\gamma(\mathbb{D}_{k})) \max\left\{r_1,r_2 \right\} \cdot \norm{\vx^\star-\vx^{p-1}}_2 \\+  \left[\eta (\sqrt{m}+2CK^2\gamma(\mathbb{D}_{k}))^2+1\right]\| \vn\|_2.
 \end{multline}

Separating $\overline{\bS^{p}\cup \bS}$ into disjoint subsets with cardinality at most $k$ and using Lemma \ref{lm:MatrixDeviationInequality}, we have
\begin{multline}\label{eq4}
    T\le 2 \max\left\{r_1,r_2 \right\} (\sqrt{m} +2CK^2\gamma(\mathbb{D}_k))\cdot \norm{\vx^\star-\vx^{p-1}}_2
    \\+2\eta  (\sqrt{m} +2CK^2\gamma(\mathbb{D}_k))^2\norm{\vn}_2.
 \end{multline}

Combining the above results \eqref{eq5:ROT}, \eqref{eq8:ROT}, \eqref{eq3} and \eqref{eq4} leads to the main results in Theorem \ref{Thm_ROT}. 
\end{proofsketch}

Next, we give the convergence analysis for ROTP algorithms.
\begin{theorem}\label{Thm_ROTP}
Let the signal model be defined as in Theorem \ref{Thm_ROT}. Let  $\eta $ be the step size in the algorithm ROTP satisfying $\eta m\le 1$. Then if the number of measurements $m$ is large enough such that $\max\{r_1,r_2\}<1$, for any sequence $\left\{ \vx^p\right\}$ generated by ROTP, one has 
\begin{equation*}\label{Re1}
    \norm{ \vx^\star-\vx^{p}}_2\le \rho_2 \norm{ \vx^\star-\vx^{p-1}}_2+c_3 \norm{\vn}_2
\end{equation*}
with probability $1-c\exp(-\gamma^2(\mathbb{D}_{2k}))$, where 
\begin{align*}
        \rho_2=&\rho_1/\sqrt{1-\left(\max\{r_1,r_2\}\right)^2}, \nonumber \\
         c_3=&\frac{c_2}{\sqrt{1-\left(\max\{r_1,r_2\}\right)^2}}+\frac{\eta   (\sqrt{m} +2CK^2\gamma(\mathbb{D}_k))}{ \left[1-\left(\max\{r_1,r_2\}\right)\right]}.
\end{align*}
Let $m$ be suitably large such that $\rho_2< 1$. Then  $\left\{ \vx^p\right\}$ can approximate $\vx^\star$ with the error
\begin{equation}\label{Re2}
    \| \vx^\star-\vx^{p}\|_2\le \rho_2^p \| \vx^\star-\vx^{0}\|_2+ \frac{c_3}{1-\rho_2}  \norm{\vn}_2
\end{equation}
with probability $1-c'\exp(-\gamma^2(\mathbb{D}_{2k}))$.  In particular, if $\vn=\bm{0}$, the sequence $\left\{ \vx^p\right\}$ generated by ROTP converges to $\vx^\star$.
\end{theorem}
\begin{proofsketch}
Since ROTP and ROT share the same steps from $3$ to $5$, we can use Theorem \ref{Thm_ROT} to bound $\|\vx^\star-\vv \|_2$ as follows
\begin{equation}\label{eq1a:ROTP}
    \norm{ \vx^\star-\vv}_2\le \rho_1 \norm{ \vx^\star-\vx^{p-1}}_2+c_2 \norm{\vn}_2.
\end{equation}

Next, we will relate  $\norm{ \vx^\star-\vx^{p}}_2$ with $\norm{ \vx^\star-\vv}_2$.  Due to the fact that $(\vx^{p})_{\overline{\mathbb{S}^{p}}}=0$ and $\vv_{\overline{\mathbb{S}^{p}}}=0$,  we have 
\begin{align}
    \norm{ \vx^\star-\vx^{p}}_2^2
    &=\norm{\left( \vx^\star-\vx^{p}\right)_{\mathbb{S}^{p}}}_2^2+\norm{\left( \vx^\star-\vx^{p}\right)_{\overline{\mathbb{S}^{p}}}}_2^2  \nonumber \\
    &=\norm{\left( \vx^\star-\vx^{p}\right)_{\mathbb{S}^{p}}}_2^2+\norm{\left( \vx^\star-\vv\right)_{\overline{\mathbb{S}^{p}}}}_2^2.
\end{align}
It follows from Lemma \ref{lm:MatrixDeviationInequality} that
\begin{multline}
    \norm{\left( \vx^\star-\vx^{p}\right)_{\mathbb{S}^{p}}}_2 \le \max\{r_1,r_2\} \norm{(\vx^\star-\vx^{p})}_2 \\ + \eta (\sqrt{m} +2CK^2\gamma(\mathbb{D}_k))\norm{\vn}_2.
\end{multline}

Using the fact that $\norm{\left( \vx^\star-\vv\right)_{\overline{\mathbb{S}^{p}}}}_2 \le \norm{ \vx^\star-\vv}_2$
and the formula of roots of a quadratic equation, we can give the upper bound of $\norm{ \vx^\star-\vx^{p}}_2$ relating with  $\norm{ \vx^\star-\vv}_2$. Together with \eqref{eq1a:ROTP}, we finish the proof.

\end{proofsketch}

The above theorems show the tradeoffs in ROT or ROTP between the number of measurements and the number of iterations. Note that if $m$ is larger than some value, the convergence rate $\rho_1$ or $\rho_{2}$ is less than $1$; if not, the algorithm might diverge. Besides, notice that the value $\rho_{1}$ or $\rho_{2}$ decreases with the increase of $m$, leading to a fast convergence rate. Therefore, we can conclude that when the number of measurements is large enough, the larger the number of measurements, the faster the convergence rate and hence the smaller number of iterations for fixed accuracy. In particular, let the step size $\eta$ be equal to ${1}/{m}$.  In this case, the convergence rate satisfying $\rho_{1}<1$ for ROT algorithm is ensured by $\sqrt{m}>C'CK^2\gamma(\mathbb{D}_{2k})$, where $C'\approx 50.943$ is the positive root of the equation $t^2-50t-48=0$. Similarly, in the  ROTP algorithm, the convergence rate satisfying $\rho_{2}<1$ is guaranteed by  $\sqrt{m}>C''CK^2\gamma(\mathbb{D}_{2k})$,  where $C''\approx 52.2614$ is the positive real root of the equation $t^4-4t^3-2444t^2-4032t-2880=0$. 

Note that in the noise-free situation, i.e., $\vn=\bm{0}$, if the conditions $\sqrt{m}>50.943 C K^2\gamma(\mathbb{D}_{2k})$ and $\eta m\le 1$ hold, the convergence rate in Theorem \ref{Thm_ROT} is strictly less than 1. It means that a linear convergence rate can be achieved by the algorithm ROT. Similarly, the algorithm ROTP also achieves a linear convergence rate when a certain condition holds.

It should be pointed out that the Gaussian complexity $\gamma(\mathbb{D}_{k})$ and the Gaussian width $w(\mathbb{D}_{k})$ have the same order due to Eq. \eqref{eq:cworder}. This, combined with Fact \ref{fact2},  tells us that $\gamma(\mathbb{D}_{k})$ is of the order of $\sqrt{k \log({n}/{k})}$, where $n$ and $k$ are the dimension and sparsity of the target signal, respectively. Based on the main results in Theorems \ref{Thm_ROT}, \ref{Thm_ROTP} and the above remarks, it is not hard to see that the number of measurements required for successful recovery is of the order of $k \log({n}/{k})$.

\begin{figure}[!t]
\centering
\begin{minipage}[b]{0.6\linewidth}
  \centering
  \centerline{\includegraphics[width=8.2cm]{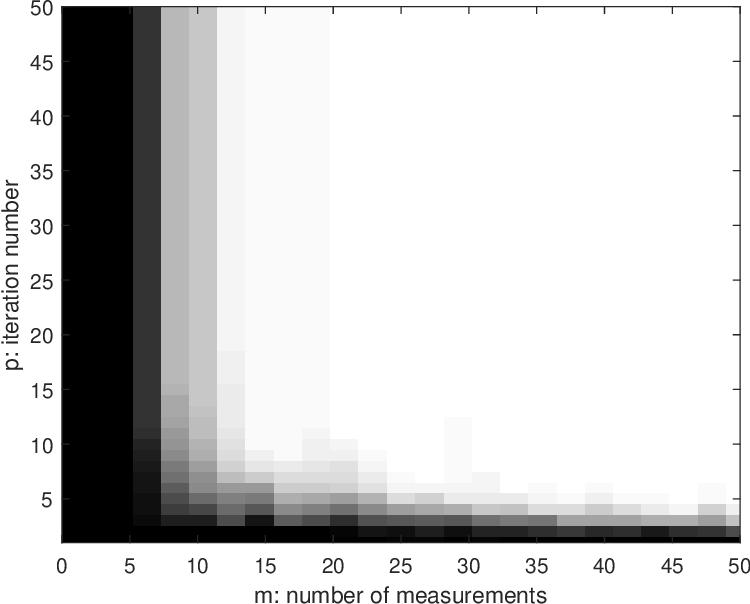}}
  \centerline{(a) ROT algorithm}\medskip
\end{minipage}

\begin{minipage}[b]{0.6\linewidth}
  \centering
  \centerline{\includegraphics[width=8.20cm]{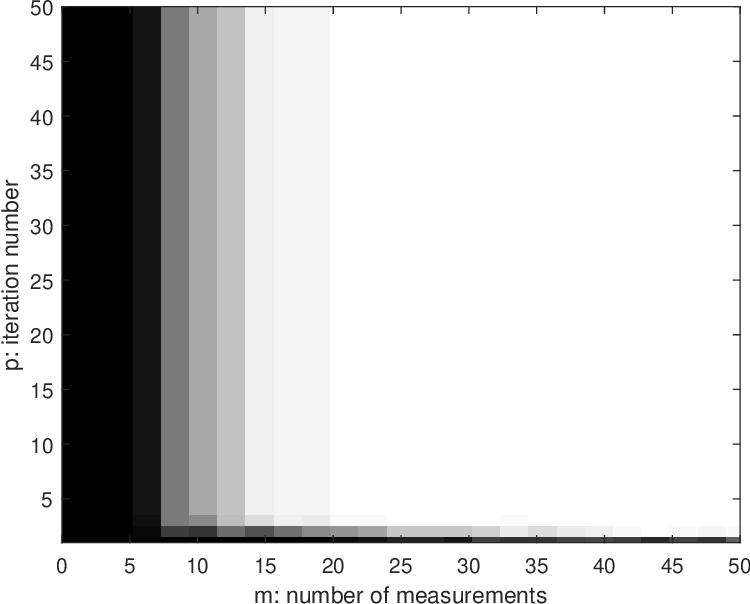}}
  \centerline{(b) ROTP algorithm}\medskip
\end{minipage}
\caption{The phase transition curves of optimal $k$-thresholding algorithms in compressed sensing: (a) ROT algorithm. (b) ROTP algorithm. The brightness of each point denotes the probability of successful reconstruction, ranging from black (0\%) to white (100\%). }
\label{fig:phasetransition}
\end{figure}

\section{Simulations}
\label{sec:Simulations}

In this section, we make simulations to validate the theory.  To show the tradeoffs between the number of measurements $m$ and the number of iterations $p$, we draw the phase transitions for ROT and ROTP algorithms. The target signal $\vx^\star$ is a $k$-sparse $n$-dimensional vector, whose nonzero entries follow standard Gaussian distribution. The sensing matrix $\vA \in \R^{m \times n}$ is a Bernoulli matrix whose entries are independent symmetric Bernoulli variables. In this simulation, $\vx^0$ is set to be $\bm{0}$, the number of dimension is set to be $n=50$, the sparsity is set to be $k=3$, the step size is set to be $\eta=1/m$, and the tolerance for stopping criteria is set to be $\varepsilon=10^{-2}$. For a pair of $m$ and $p$, we run $50$ Monte Carlo trials and count the number of successful trials which satisfy  ${\norm{\vx^\star-\vx^{p}}_2}/{\norm{\vx^\star}_2} \le \varepsilon$. Then we calculate the successful probability for each pair and get the phase transition results.

The numerical results are shown in Fig. \ref{fig:phasetransition}. Note the white area for ROTP algorithm is larger than that for ROT algorithm, which means ROTP algorithm has a better recovery performance than ROT algorithm. Besides, when the number of measurements is larger than 20, ROTP needs less iterations to achieve convergence. Both algorithms illustrate the data-time tradeoffs as discussed in Section \ref{sec:TheoreticalGuarantees}. The curves present that when the number of measurements is smaller than a certain number, the algorithms cannot recover the target signal. Moreover, when the number of measurements is larger than a certain number, the number of iterations required for successful recovery decreases with the increase of the the number of measurements. The observed phenomenon in the simulations corresponds with our theoretical results in Theorems \ref{Thm_ROT} and \ref{Thm_ROTP}. 


\section{Conclusion and future work}
\label{sec:Conclusion}

In this paper, we have provided the analysis for the data-time tradeoffs for optimal $k$-thresholding algorithms in compressed sensing. The theory has shown that the number of measurements should be suitably large  to guarantee the convergence of the optimal $k$-thresholding algorithms. It also presented that the number of measurements required for successful reconstruction has a negative correlation with the number of iterations. Simulations have been provided to validate the correctness of our theory.

For future work, it would be interesting to extend the data-time tradeoffs analysis for optimal $k$-thresholding algorithms from compressed sensing to other important linear inverse problems such as quantized compressed sensing, corrupted sensing and so on. 

\clearpage

\clearpage
\onecolumn

\setcounter{equation}{0}
\renewcommand{\theequation}{A.\arabic{equation}}

\setcounter{theorem}{0}
\renewcommand{\thetheorem}{A\arabic{theorem}}

\setcounter{lemma}{0}
\renewcommand{\thelemma}{A.\arabic{lemma}}

\setcounter{fact}{0}
\renewcommand{\thefact}{A.\arabic{fact}}

\appendices

\section{Proof of Theorem \ref{Thm_ROT}}

Before proving the theorem, we review some important facts.
\begin{fact}[\cite{zhao2020optimal}]\label{lm:ROTPPROOF2}
Let $\vx$ be a $k$-sparse vector and $\mathbb{S}$ be the support of $\vx$. Let $\vw^p$ be a solution to the problem \eqref{HT-QP-Relax}, and $\vv={\cal H}_k(\vu^p \odot \vw^p)$ be a vector with the support $\mathbb{S}^{p+1}$. Then we have 
$$\|\vx-\vv\|_2\leq \|(\vu^p \odot \vw^p-\vx)_{\mathbb{S}^{p+1}\cup \mathbb{S}}\|_2+\| (\vu^p \odot \vw^p-\vx)_{\mathbb{S}^{p+1}\setminus \mathbb{S}}\|_2$$
\end{fact}

\begin{fact}[\cite{zhao2020optimal}] \label{lm:ROTPPROOF3}
Let $k$ and $\overline{k}$ be two integer numbers such that $0\le \overline{k}\le k$. Let $\vw\in \mathcal{W}^k=\left\{\vw:\ve^T\vw=k,~\vw\in\left\{0,1\right\}^n  \right\}$ be the vector with the support $\mathbb{S}$, and $\overline{\bS}$ be the complementary set. Let $\overline{\bS}$ has the disjointed partition $\bS_i,~i=1,...,\overline{n}$ such that 
$$\overline{\bS}=\bS_1\cup \bS_2\cup...\cup \bS_{\overline{n}-1}\cup \bS_{\overline{n}}$$
satisfying the following properties:
\begin{itemize}
    \item $\vert \bS_i\vert\le k$ for $i=1,2,...,\overline{n}-1$ and $\vert \bS_{\overline{n}}\vert\le \overline{k}<k$;
    \item $\bS_i$ is the index set for the $i$th largest $k$ components in $\vw_{\overline{\bS}}$.  
\end{itemize}
Let $\alpha_{(i)}$ denote the largest entry of $\vw$ on the support $\bS_i$. Then we have
$$\sum_{i=1}^{\overline{n}}\alpha_{(i)}\le 2-\frac{1}{k}<2.$$
\end{fact} 

Now we are ready to prove the theorem. By using $\vu^p=\vx^p+\eta \vA^T(\vy-\vA \vx^p)$ and $\vy=\vA\vx^\star+\vn$, we obtain
\begin{equation}\label{transaction}
    \vx^\star-\vu^p  =(\vI - \eta  \vA^T \vA) (\vx^\star-\vx^p)-\eta  \vA^T \vn.
\end{equation}
Let $\vw(\vu^p)$ be a solution to the following problem in ROT at the $p$-th iteration
 \begin{equation}   \label {HT-QP-Relax}  \min_{\vw}   \left\{  \|\vy- \vA (\vu^p  \odot   \vw) \|^2_2 : ~  \textrm{\textbf{e}}^T \vw= k , ~  \bm{0}\leq w\leq \textrm{\textbf{e}} \right\}.
  \end{equation}
Then the next iteration is updated by $\vx^{p+1}\in  {\cal H}_k (\vu^p  \odot  \vw(\vu^p))$. Let $\bS^{p+1}$ and $\bS$ be the support of $\vx^{p+1}$ and $\vx^\star$, respectively. By Fact \ref{lm:ROTPPROOF2}, we have 
\begin{equation}\label{eq1:ROT}
\|\vx^\star-\vx^{p+1}\|_2\leq \| \left[\vu^p \odot \vw(\vu^p)-\vx^\star\right]_{\bS^{p+1}\setminus \bS}\|_2+\|\left[\vu^p \odot \vw(\vu^p)-\vx^\star\right]_{\bS^{p+1}\cup \bS}\|_2.    
\end{equation}
Now we estimate the first term of the right side of \eqref{eq1:ROT}. Due to the fact that $\vx^\star_{\bS^{p+1}\setminus \bS}=0$ and $\vw_i^p\in [0,1],i=1,\ldots,n$, we have
\begin{align}  
    \| \left[\vu^p \odot \vw(\vu^p\right]-\vx^\star)_{\bS^{p+1}\setminus \bS}\|_2
    &= \| \left[\vu^p \odot \vw(\vu^p)\right]_{\bS^{p+1}\setminus \bS}\|_2 =\| \left[(\vu^p-\vx^\star) \odot \vw(\vu^p)\right]_{\bS^{p+1}\setminus \bS}\|_2 \nonumber\\
    &=  \| \left[(\eta  \vA^T \vn-(\vI - \eta  \vA^T \vA) (\vx^\star-\vx^p)) \odot \vw(\vu^p)\right]_{\bS^{p+1}\setminus \bS}\|_2 \nonumber\\
    &\le \eta \| \left[  \vA^T \vn  \right]_{\bS^{p+1}\setminus \bS}\|_2+ \| \left[(\vI - \eta  \vA^T \vA) (\vx^\star-\vx^p) \right]_{\bS^{p+1}\setminus \bS}\|_2,\label{eq2:ROT}
\end{align}
where the third equality follows from \eqref{transaction}. By using Lemma \ref{lm:OTROT} with $\vv=\vx^\star-\vx^p\in \mathbb{D}_{2k}$ and $\bL=\bS^{p+1}\setminus \bS $, we can estimate the two terms in the above inequality. Note that $\mathbb{D}_l=\mathbb{D}_k$. Then we have 
\begin{equation}\label{eq3:ROT}
    \norm{[(\vI-\eta \vA^T \vA)(\vx^\star-\vx^p)]_{\bS^{p+1}\setminus \bS}}_2 \le \max\left\{r_1,r_2 \right\} \cdot \norm{\vx^\star-\vx^p}_2
\end{equation}
and 
\begin{equation}\label{eq4:ROT}
    \norm{[\vA^T \vn]_{\bS^{p+1}\setminus \bS}}_2 \le  \norm{\vn}_2(\sqrt{m} +2CK^2\gamma(\mathbb{D}_k))
\end{equation}
where
$$r_1=1-\eta m \cdot    \frac{\sqrt{m}-6CK^2\gamma(\mathbb{D}_{2k}+\mathbb{D}_k)}{\sqrt{m}},~~
        r_2=\frac{3CK^2\gamma(\mathbb{D}_{2k}+\mathbb{D}_k)}{\sqrt{m}}\left(\frac{3CK^2\gamma(\mathbb{D}_{2k}+\mathbb{D}_k)}{2\sqrt{m}}+1\right).$$
Note that $\gamma(\mathbb{D}_{2k}+\mathbb{D}_{k})\le \gamma(\mathbb{D}_{2k}+\mathbb{D}_{2k})\le 2\gamma(\mathbb{D}_{2k})$ which follows from \textbf{b)} and \textbf{c)} in Lemma \ref{lm:Dk}.
The constants $r_1$ and $r_2$ can be simplified to 
\begin{equation}\label{rho}
    r_1=1-\eta m \cdot    \frac{\sqrt{m}-12CK^2\gamma(\mathbb{D}_{2k})}{\sqrt{m}},~~
        r_2=\frac{6CK^2\gamma(\mathbb{D}_{2k})}{\sqrt{m}}\left(\frac{3CK^2\gamma(\mathbb{D}_{2k})}{\sqrt{m}}+1\right).
\end{equation}
Combining \eqref{eq2:ROT}, \eqref{eq3:ROT} and \eqref{eq4:ROT} yields that 
\begin{equation}\label{eq5:ROT}
    \| \left[\vu^p \odot \vw(\vu^p)-\vx^\star\right]_{\bS^{p+1}\setminus \bS}\|_2\le \max\left\{r_1,r_2 \right\} \cdot \norm{\vx^\star-\vx^p}_2+\eta \norm{\vn}_2(\sqrt{m} +2CK^2\gamma(\mathbb{D}_k)).
\end{equation}
Next, we estimate the second term of the right side of \eqref{eq1:ROT}. By using Lemma \ref{lm:MatrixDeviationInequality} with $\mathbb{D}=\mathbb{D}_{2k}$, we have
\begin{equation}\label{eq6:ROT}
    \norm{ \vA\frac{\left[\vu^p \odot \vw(\vu^p)-\vx^\star\right]_{\bS^{p+1}\cup \bS}}{\| \left[\vu^p \odot \vw(u^p)-\vx^\star\right]_{\bS^{p+1}\cup \bS}\|_2} }_2\ge \sqrt{m}-2CK^2\gamma(\mathbb{D}_{2k})
\end{equation}
with probability at least $1-2\mathrm{exp}(-\gamma^2(\mathbb{D}_{2k}))$. 
Note that 
\begin{align}  \label{eq7:ROT}
    \| y-\vA\left[\vu^p \odot \vw(\vu^p)\right]\|_2
    &= \| \vA\left[\vu^p \odot \vw(\vu^p)-\vx^\star\right]-\vn\|_2 \nonumber\\ 
    &= \|  \vA\left[\vu^p \odot \vw(\vu^p)-\vx^\star\right]_{\bS^{p+1}\cup \bS}+ \vA\left[\vu^p \odot \vw(\vu^p)-\vx^\star\right]_{\overline{\bS^{p+1}\cup \bS}}-\vn\|_2 \nonumber\\
    &\ge \|  \vA\left[\vu^p \odot \vw(\vu^p)-\vx^\star\right]_{\bS^{p+1}\cup \bS}\|_2 - \|  \vA\left[\vu^p \odot \vw(\vu^p)-\vx^\star\right]_{\overline{\bS^{p+1}\cup \bS}}\|_2-\|\vn\|_2 \nonumber\\
    &\ge \| \left[\vu^p \odot \vw(\vu^p)-\vx^\star\right]_{\bS^{p+1}\cup \bS}\|_2 (\sqrt{m}-2CK^2\gamma(\mathbb{D}_{2k})) - \|  \vA\left[\vu^p \odot \vw(\vu^p)-\vx^\star\right]_{\overline{\bS^{p+1}\cup \bS}}\|_2-\|\vn\|_2,
\end{align}
where the first equality uses $y=\vA\vx^\star+\vn$, the first inequality uses the triangle inequality and the last inequality follows from \eqref{eq6:ROT}. Thus we have 
\begin{equation}\label{eq8:ROT}
    \| \left[\vu^p \odot \vw(\vu^p)-\vx^\star\right]_{\bS^{p+1}\cup \bS}\|_2\le \frac{1}{\sqrt{m}-2CK^2\gamma(\mathbb{D}_{2k})} \left\{ \| y-\vA\left[\vu^p \odot \vw(\vu^p)\right]\|_2+T+\|\vn\|_2 \right\},
\end{equation}
where $T=\|  \vA\left[\vu^p \odot \vw(\vu^p)-\vx^\star\right]_{\overline{\bS^{p+1}\cup \bS}}\|_2$.
Let $\hat{\vw}\in \mathcal{W}^{(k)}$ be the vector such that $\vx^\star=\vx^\star \odot \hat{\vw}.$
\begin{align}  \label{eq9:ROT}
    \| y-\vA\left[\vu^p \odot \vw(\vu^p)\right]\|_2
    &\le  \| y-\vA(\vu^p \odot \hat{\vw})\|_2, \nonumber\\ 
     & = \| \vA\vx^\star+\vn-\vA(\vu^p \odot \hat{\vw})\|_2= \|  [\vA(\vx^\star-\vu^p) \odot \hat{\vw}]+\vn\|_2, \nonumber\\
      &\le  \|  \vA \left[(\vx^\star-\vu^p) \odot \hat{\vw}\right]\|_2+ \|  \vn\|_2, \nonumber\\
    &\le (\sqrt{m}+2CK^2\gamma(\mathbb{D}_{k}))\|  (\vx^\star-\vu^p) \odot \hat{\vw}\|_2+ \|  \vn\|_2,
\end{align}
where the second inequality uses the triangle inequality and the last inequality follows from Lemma \ref{lm:MatrixDeviationInequality} with $\mathbb{D}=\mathbb{D}_{k}$.
Using the fact that $ \vx^\star-\vu^p  =(\vI - \eta  \vA^T \vA) (\vx^\star-\vx^p)-\eta  \vA^T \vn$ and Lemma \ref{lm:OTROT} yields
\begin{align}  \label{eq10:ROT}
    \|  (\vx^\star-\vu^p) \odot \hat{\vw}\|_2
    &\le  \norm{\[(\vI - \eta  \vA^T \vA) (\vx^\star-\vx^p)\] \odot \hat{\vw}}_2+ \eta \norm { \[\vA^T \vn\]\odot \hat{\vw}}_2, \nonumber\\
      &\le  \max\left\{r_1,r_2 \right\} \cdot \norm{\vx^\star-\vx^p}_2+\eta (\sqrt{m} +2CK^2\gamma(\mathbb{D}_k))\norm{\vn}_2,
\end{align}
 where $r_1$ and $r_2$ are given in \eqref{rho}.
Combining \eqref{eq9:ROT} and \eqref{eq10:ROT} leads to 
\begin{equation}\label{eq11:ROT}
      \| y-\vA\left[\vu^p \odot \vw(\vu^p)\right]\|_2\le (\sqrt{m}+2CK^2\gamma(\mathbb{D}_{k})) \max\left\{r_1,r_2 \right\} \cdot \norm{\vx^\star-\vx^p}_2+  \left[\eta (\sqrt{m}+2CK^2\gamma(\mathbb{D}_{k}))^2+1\right]\| \vn\|_2.
\end{equation}
In the last part of the proof, we estimate the term $T$ in  \eqref{eq8:ROT}. Due to the  fact that $\vx^{*}_{\overline{S^{p+1}\cup S}}=0$, $T$ has the following form
$$T=\|  \vA\left[\vu^p \odot \vw(\vu^p)-\vx^\star\right]_{\overline{\bS^{p+1}\cup \bS}}\|_2=\|\vA\left[(\vu^p-\vx^\star) \odot \vw(\vu^p)\right]_{\overline{\bS^{p+1}\cup \bS}}\|_2.$$
      Let $\bT_i, ~i=1,2,...,\overline{n}$ be the disjoint subset of $\overline{\bS^{p+1}\cup \bS}$ such that
      $$\overline{\bS^{p+1}\cup \bS}=\bT_1\cup \bT_2\cup...\cup \bT_{\overline{n}}$$
        and satisfying the following properties
        \begin{itemize}
            \item $\vert \bT_i\vert=k$ for all $i=1,2,...,\overline{n}$ and $\vert \bT_{\overline{n}}\vert=\overline{k}<k$;
            \item $\bT_i$ is the index set for $i$th largest $k$ components in $\vw(\vu^p)_{\overline{\bS^{p+1}\cup \bS}}$.
        \end{itemize}
It is obvious that $\vert \overline{\bS^{p+1}\cup \bS}\vert=(\overline{n}-1)k+ \overline{k}$. Based on such partition of $\overline{\bS^{p+1}\cup \bS}$, the term $\left[(\vu^p-\vx^\star) \odot \vw(\vu^p)\right]_{\overline{\bS^{p+1}\cup \bS}}$ can be written as
$$\left[(\vu^p-\vx^\star) \odot \vw(\vu^p)\right]_{\overline{\bS^{p+1}\cup \bS}}=\left[(\vu^p-\vx^\star) \odot \vw(\vu^p)\right]_{\bT_1}+\left[(\vu^p-\vx^\star) \odot \vw(\vu^p)\right]_{\bT_2}+...+\left[(\vu^p-\vx^\star) \odot \vw(\vu^p)\right]_{\bT_{\overline{n}}}.$$ Thus $T$ can be reduced to 
\begin{align}  \label{eq12:ROT}
    T&=\|\sum_{i}^{\overline{n}} \vA\left[(\vu^p-\vx^\star) \odot \vw(\vu^p)\right]_{\bT_i}\|_2\le \sum_{i}^{\overline{n}}\| \vA\left[(\vu^p-\vx^\star) \odot \vw(\vu^p)\right]_{\bT_i}\|_2, \nonumber\\
    &\le  (\sqrt{m}+2CK^2\gamma(\mathbb{D}_{k}))\sum_{i}^{\overline{n}}\| \left[(\vu^p-\vx^\star) \odot \vw(\vu^p)\right]_{\bT_i}\|_2,
\end{align}
where the last inequality follows from Lemma \ref{lm:MatrixDeviationInequality}. By using the Eq. \eqref{transaction} and the triangle inequality, for each term $\| [(\vu^p-\vx^\star) \odot \vw(\vu^p)]_{\bT_i}\|_2$, we have 
\begin{align*} 
  \| \left[(\vu^p-\vx^\star) \odot \vw(\vu^p)\right]_{\bT_i}\|_2&=\| \left[ (\vI - \eta  \vA^T \vA) (\vx^\star-\vx^p)\odot \vw(\vu^p)-\eta  \vA^T \vn\odot \vw(\vu^p)\right]_{\bT_i}\|_2\\
  &\le \| \left[ (\vI - \eta  \vA^T \vA) (\vx^\star-\vx^p)\odot \vw(\vu^p)\right]_{\bT_i}\|_2+\eta\| \left[   \vA^T \vn\odot \vw(\vu^p)\right]_{\bT_i}\|_2.
\end{align*}
Define $\alpha_{i}\triangleq\max\left\{ (\vw(\vu^p))_j:~j\in \bT_i\right\}$, then 
\begin{align*} 
  \| \left[(\vu^p-\vx^\star) \odot \vw(\vu^p)\right]_{\bT_i}\|_2&\le \alpha_i\| \left[ (\vI - \eta  \vA^T \vA) (\vx^\star-\vx^p)\right]_{\bT_i}\|_2+\eta\alpha_i\| \left[   \vA^T \vn\right]_{\bT_i}\|_2,\\
  & \le \alpha_i \max\left\{r_1,r_2 \right\} \cdot \norm{\vx^\star-\vx^p}_2+\eta \alpha_i (\sqrt{m} +2CK^2\gamma(\mathbb{D}_k))\norm{\vn}_2,
\end{align*}
where $r_1$ and $r_2$ are given in \eqref{eq10:ROT}, and thus
$$\sum_{i}^{\overline{n}}\| \left[(\vu^p-\vx^\star) \odot \vw(\vu^p)\right]_{\bT_i}\|_2\le \sum_{i}^{\overline{n}}\alpha_i \max\left\{r_1,r_2 \right\} \cdot \norm{\vx^\star-\vx^p}_2+\eta \sum_{i}^{\overline{n}}\alpha_i (\sqrt{m} +2CK^2\gamma(\mathbb{D}_k))\norm{\vn}_2.$$
Using Fact \ref{lm:ROTPPROOF3} with $\vw=\vw(\vu^p),~\bS=\bS^{p+1}\cup \bS$ and $\bS_i=\bT_i$ yields
\begin{equation}\label{eq13:ROT}
    T\le 2 \max\left\{r_1,r_2 \right\} (\sqrt{m} +2CK^2\gamma(\mathbb{D}_k))\cdot \norm{\vx^\star-\vx^p}_2+2\eta  (\sqrt{m} +2CK^2\gamma(\mathbb{D}_k))^2\norm{\vn}_2.
\end{equation}
Merging \eqref{eq13:ROT}, \eqref{eq8:ROT} and \eqref{eq11:ROT} leads to 
\begin{align*} 
  \| \left[\vu^p \odot \vw(\vu^p)-\vx^\star\right]_{\bS^{p+1}\cup \bS}\|_2\le \|\left[(\vu^p-\vx^\star) \odot \vw(\vu^p)\right]_{\bS^{p+1}\cup \bS}\|_2&\le      \frac{3\max\left\{r_1,r_2 \right\}(\sqrt{m}+2CK^2\gamma(\mathbb{D}_{k}))}{\sqrt{m}-2CK^2\gamma(\mathbb{D}_{2k})}  \norm{\vx^\star-\vx^p}_2\\
  &~~~~+\frac{3\eta (\sqrt{m}+2CK^2\gamma(\mathbb{D}_{k}))^2+2}{\sqrt{m}-2CK^2\gamma(\mathbb{D}_{2k})}\norm{\vn}_2, 
\end{align*}
which combined with \eqref{eq1:ROT} and \eqref{eq5:ROT} leads to
\begin{equation}\label{eq14:ROT}
    \| \vx^\star-\vx^{p+1}\|_2\le c_1 \max\left\{r_1,r_2 \right\} \| \vx^\star-\vx^{p}\|_2+c_2 \norm{\vn}_2,
\end{equation}
where 
\begin{align*}
    c_1&=\frac{4\sqrt{m}+6CK^2\gamma(\mathbb{D}_{k})-2CK^2\gamma(\mathbb{D}_{2k})}{\sqrt{m}-2CK^2\gamma(\mathbb{D}_{2k})} \le \frac{4\sqrt{m}+4CK^2\gamma(\mathbb{D}_{k})}{\sqrt{m}-2CK^2\gamma(\mathbb{D}_{2k})},\\
    c_2&=\frac{3\eta (\sqrt{m}+2CK^2\gamma(\mathbb{D}_{k}))^2+2}{\sqrt{m}-2CK^2\gamma(\mathbb{D}_{2k})}+\eta(\sqrt{m}+2CK^2\gamma(\mathbb{D}_{k})).
\end{align*}
Let $ \rho_1= c_1 \max\left\{r_1,r_2 \right\}$. By using the fact that $\sum_{j=1}^{\infty}s^j=\frac{1}{1-s}$ when $0<s<1$, the result \eqref{eq14:ROT} can be reduced to 
$$ \| \vx^\star-\vx^{p+1}\|_2\le \rho_1^{p+1} \| \vx^\star-\vx^{0}\|_2+ \frac{c_2}{1-\rho_1}  \norm{\vn}_2,$$
as long as $0<\rho_1<1$.

\section{Proof of Theorem \ref{Thm_ROTP}}
Since ROTP and ROT share the same steps from $3$ to $5$, we can use Theorem \ref{Thm_ROT} to bound $\|\vx^\star-\vv \|_2$ as follows
\begin{equation}\label{eq1:ROTP}
    \| \vx^\star-\vv\|_2\le \rho_1 \| \vx^\star-\vx^{p}\|_2+c_2 \norm{\vn}_2.
\end{equation}
Next, we will relate  $\| \vx^\star-\vx^{p+1}\|_2$ with $\| \vx^\star-\vv\|_2$. Due to $\vx^{p+1}$ is the solution of the problem
$$\vx^{p+1}= \mathrm{argmin}_{\vx} \big{\{}\|\vy- \vA \vx \|^2_2, ~ \mbox{supp}(\vx) \subseteq \mathbb{S}^{p+1} \big{\}},$$
then we have $\left[\vA^T(\vy-\vA \vx^{p+1}) \right]_{\mathbb{S}^{p+1}}=0$. Together with $\vy=\vA\vx^\star+\vn$, we have
\begin{equation}
    \left[(\vI-\eta \vA^T\vA) (\vx^\star-\vx^{p+1})\right]_{\mathbb{S}^{p+1}}= \left( \vx^\star-\vx^{p+1}\right)_{\mathbb{S}^{p+1}}+\eta \left(\vA^T\vn \right)_{\mathbb{S}^{p+1}}.
\end{equation}
By using the triangle inequality, we obtain
\begin{equation}
    \norm{\left( \vx^\star-\vx^{p+1}\right)_{\mathbb{S}^{p+1}}}_2 \le \norm{\left[(\vI-\eta \vA^T\vA) (\vx^\star-\vx^{p+1})\right]_{\mathbb{S}^{p+1}}}_2+ \eta \norm{\left(\vA^T\vn \right)_{\mathbb{S}^{p+1}}}_2.
\end{equation}
Since the cardinality of $\mathbb{S}^{p+1}$ is $k$ and $(\vx^\star-\vx^{p+1})$ is a $2k$-sparse vector, using Lemma \ref{lm:OTROT} yields
\begin{equation}
    \norm{\left[(\vI-\eta \vA^T\vA) (\vx^\star-\vx^{p+1})\right]_{\mathbb{S}^{p+1}}}_2 \le \max\{r_1,r_2\} \norm{(\vx^\star-\vx^{p+1})}_2,
\end{equation}
and 
\begin{equation}
    \norm{\left(\vA^T\vn \right)_{\mathbb{S}^{p+1}}}_2 \le (\sqrt{m} +2CK^2\gamma(\mathbb{D}_k))\norm{\vn}_2.
\end{equation}
So we get
\begin{equation}
    \norm{\left( \vx^\star-\vx^{p+1}\right)_{\mathbb{S}^{p+1}}}_2 \le \max\{r_1,r_2\} \norm{(\vx^\star-\vx^{p+1})}_2+ \eta (\sqrt{m} +2CK^2\gamma(\mathbb{D}_k))\norm{\vn}_2.
\end{equation}
Due to the fact that $(\vx^{p+1})_{\overline{\mathbb{S}^{p+1}}}=0$ and $\vv_{\overline{\mathbb{S}^{p+1}}}=0$,  we have 
\begin{align}
    \norm{ \vx^\star-\vx^{p+1}}_2^2
    &=\norm{\left( \vx^\star-\vx^{p+1}\right)_{\mathbb{S}^{p+1}}}_2^2+\norm{\left( \vx^\star-\vx^{p+1}\right)_{\overline{\mathbb{S}^{p+1}}}}_2^2=\norm{\left( \vx^\star-\vx^{p+1}\right)_{\mathbb{S}^{p+1}}}_2^2+\norm{\left( \vx^\star-\vv\right)_{\overline{\mathbb{S}^{p+1}}}}_2^2  \nonumber \\
    &\le \left(\max\{r_1,r_2\}\right)^2 \norm{(\vx^\star-\vx^{p+1})}_2^2+ \eta^2 (\sqrt{m} +2CK^2\gamma(\mathbb{D}_k))^2\norm{\vn}_2^2 \nonumber \\ 
    &\quad +2 \eta \max\{r_1,r_2\}  (\sqrt{m} +2CK^2\gamma(\mathbb{D}_k))\norm{(\vx^\star-\vx^{p+1})}_2 \norm{\vn}_2+\norm{\left( \vx^\star-\vv\right)_{\overline{\mathbb{S}^{p+1}}}}_2^2,
\end{align}
which is equivalent to 
\begin{multline}
    \left[1-\left(\max\{r_1,r_2\}\right)^2\right] \norm{ \vx^\star-\vx^{p+1}}_2^2  \\
    \le \eta^2 (\sqrt{m} +2CK^2\gamma(\mathbb{D}_k))^2\norm{\vn}_2^2
     +2 \eta \max\{r_1,r_2\}  (\sqrt{m} +2CK^2\gamma(\mathbb{D}_k))\norm{(\vx^\star-\vx^{p+1})}_2 \norm{\vn}_2+\norm{\left( \vx^\star-\vv\right)_{\overline{\mathbb{S}^{p+1}}}}_2^2.
\end{multline}

By using the formula of roots of a quadratic equation, we can give the upper bound of $\norm{ \vx^\star-\vx^{p+1}}_2$
\begin{align}
    &  \norm{ \vx^\star-\vx^{p+1}}_2 \nonumber \\
    & \le \frac{2 \eta \max\{r_1,r_2\}  (\sqrt{m} +2CK^2\gamma(\mathbb{D}_k)) \norm{\vn}_2 + \sqrt{4\eta^2  (\sqrt{m} +2CK^2\gamma(\mathbb{D}_k))^2 \norm{\vn}_2^2+4\left[1-\left(\max\{r_1,r_2\}\right)^2\right] \norm{\left( \vx^\star-\vv\right)_{\overline{\mathbb{S}^{p+1}}}}_2^2}}{2 \left[1-\left(\max\{r_1,r_2\}\right)^2\right]} \nonumber \\
    &\le \frac{2 \eta \max\{r_1,r_2\}  (\sqrt{m} +2CK^2\gamma(\mathbb{D}_k)) \norm{\vn}_2 + 2\eta  (\sqrt{m} +2CK^2\gamma(\mathbb{D}_k)) \norm{\vn}_2+2\sqrt{1-\left(\max\{r_1,r_2\}\right)^2} \norm{\left( \vx^\star-\vv\right)_{\overline{\mathbb{S}^{p+1}}}}_2}{2 \left[1-\left(\max\{r_1,r_2\}\right)^2\right]} \nonumber \\
    &=\frac{\eta   (\sqrt{m} +2CK^2\gamma(\mathbb{D}_k)) \norm{\vn}_2}{ \left[1-\left(\max\{r_1,r_2\}\right)\right]}
    +\frac{ \norm{\left( \vx^\star-\vv\right)_{\overline{\mathbb{S}^{p+1}}}}_2}{\sqrt{1-\left(\max\{r_1,r_2\}\right)^2}}\le \frac{\eta   (\sqrt{m} +2CK^2\gamma(\mathbb{D}_k)) \norm{\vn}_2}{ \left[1-\left(\max\{r_1,r_2\}\right)\right]}
    +\frac{ \norm{ \vx^\star-\vv}_2}{\sqrt{1-\left(\max\{r_1,r_2\}\right)^2}},
\end{align}
where the second inequality uses $a^2+b^2 \le (a+b)^2$ for any $a,b\ge 0$ and the last inequality follows from the fact $\norm{\left( \vx^\star-\vv\right)_{\overline{\mathbb{S}^{p+1}}}}_2\le \norm{ \vx^\star-\vv}_2$.

Together with (\ref{eq1:ROTP}), we complete the proof
\begin{align}
      \norm{ \vx^\star-\vx^{p+1}}_2 
    \le \frac{\eta   (\sqrt{m} +2CK^2\gamma(\mathbb{D}_k)) \norm{\vn}_2}{ \left[1-\left(\max\{r_1,r_2\}\right)\right]}
    +\frac{\rho_1 \| \vx^\star-\vx^{p}\|_2+c_2 \norm{\vn}_2 }{\sqrt{1-\left(\max\{r_1,r_2\}\right)^2}}\le \rho_2 \| \vx^\star-\vx^{p}\|_2 + c_3 \norm{\vn}_2
\end{align}
where
\begin{align}
    \rho_2=\frac{\rho_1}{\sqrt{1-\left(\max\{r_1,r_2\}\right)^2}},~c_3=\frac{c_2}{\sqrt{1-\left(\max\{r_1,r_2\}\right)^2}}+\frac{\eta   (\sqrt{m} +2CK^2\gamma(\mathbb{D}_k))}{ \left[1-\left(\max\{r_1,r_2\}\right)\right]}.
\end{align}
Let $\rho_2=\rho_1/\sqrt{1-\left(\max\{r_1,r_2\}\right)^2}$. Similar to the proof of Theorem \ref{Thm_ROT}, we obtain the desired results \eqref{Re2}. 

\section{Proof of Lemmas}

In this section, we will provide the proof for the lemmas.

\subsection{Proof of Lemma \ref{lm:OTROT}}
\textbf{Proof of part a)}:
Let
\begin{equation}
    \bm{\mu}=\frac{[(\vI-\eta \vA^T \vA)\vv]_\mathbb{L} }{\norm{[(\vI - \eta  \vA^T \vA) \vv]_\mathbb{L} }_2} \in \mathbb{D}_l~~ \mathrm{and}~~ \bm{\nu}=\frac{\vv}{\norm{\vv}_2} \in \mathbb{D}_k.
\end{equation}
So we have
\begin{align}
    \norm{[(\vI - \eta  \vA^T \vA) \vv]_\mathbb{L} }_2&=
    \ip{[(\vI-\eta \vA^T \vA)\vv]_\mathbb{L} }{\bm{\mu}}=\ip{(\vI -\eta  \vA^T \vA) \vv }{\bm{\mu}}=\norm{\vv}_2\ip{(\vI - \eta \vA^T \vA)  \bm{\nu} }{\bm{\mu}}.
\end{align}
As for $\ip{(\vI - \eta  \vA^T \vA) \bm{\nu}}{\bm{\mu}}$, we have 
\begin{align}
    \ip{(\vI - \eta  \vA^T \vA) \bm{\nu}}{\bm{\mu}}&= \ip{ \bm{\nu}}{\bm{\mu}}-\eta  \ip{\vA \bm{\nu}}{\vA\bm{\mu}}=\frac{1}{4}\left(\norm{ \bm{\nu}+\bm{\mu}}_2^2 -\eta  \norm{\vA( \bm{\nu}+\bm{\mu})}_2^2-\norm{ \bm{\nu}-\bm{\mu}}_2^2+\eta \norm{\vA( \bm{\nu}-\bm{\mu})}_2^2\right).
\end{align}
Notice that $\mathbb{D}_l=-\mathbb{D}_l$, so we have 
\begin{align}
     \bm{\nu}+\bm{\mu} \in \mathbb{D}_k+\mathbb{D}_l, ~ \bm{\nu}-\bm{\mu} \in \mathbb{D}_k-\mathbb{D}_l=\mathbb{D}_k+\mathbb{D}_l.
\end{align}
 Using Fact \ref{lm:MatrixDeviationInequality} with $\mathbb{D}=\mathbb{D}_k+\mathbb{D}_l$, $t=\gamma(\mathbb{D}_k+\mathbb{D}_l)$ and $ \mbox{rad}(\mathbb{D}_k+\mathbb{D}_l)\leq 2$, we obtain 
\begin{equation}
    \norm{\vA( \bm{\nu}+\bm{\mu})}_2^2 \ge \left(\max\{\sqrt{m} \norm{ \bm{\nu}+\bm{\mu}}_2 -3CK^2\gamma(\mathbb{D}_k+\mathbb{D}_l),0\}\right)^2
\end{equation}
and
\begin{equation}
    \norm{\vA( \bm{\nu}-\bm{\mu})}_2^2 \le \left[\sqrt{m}\norm{ \bm{\nu}-\bm{\mu}}_2 +3CK^2\gamma(\mathbb{D}_k+\mathbb{D}_l)\right]^2
\end{equation}
with both probabilities at least $1-2\exp(-\gamma^2(\mathbb{D}_k+\mathbb{D}_l))$.

Next, we separate the problems into two cases.
\noindent \textbf{Case I:} If $\sqrt{m} \norm{ \bm{\nu}+\bm{\mu}}_2-3CK^2\gamma(\mathbb{D}_k+\mathbb{D}_l)\ge 0$, we have
\begin{align}
    &~~~\ip{(\vI - \eta  \vA^T \vA) \bm{\nu}}{\bm{\mu}}= \ip{ \bm{\nu}}{\bm{\mu}}-\eta  \ip{\vA \bm{\nu}}{\vA\bm{\mu}} \nonumber\\
    &=\frac{1}{4}\left[\norm{ \bm{\nu}+\bm{\mu}}_2^2 -\eta  \norm{\vA( \bm{\nu}+\bm{\mu})}_2^2-\norm{ \bm{\nu}-\bm{\mu}}_2^2+\eta  \norm{\vA( \bm{\nu}-\bm{\mu})}_2^2\right) \nonumber\\
    &\le \frac{1}{4}\left[\norm{ \bm{\nu}+\bm{\mu}}_2^2 -\eta \left(\sqrt{m} \norm{ \bm{\nu}+\bm{\mu}}_2 -3CK^2\gamma(\mathbb{D}_k+\mathbb{D}_l)\right)^2-\norm{ \bm{\nu}-\bm{\mu}}_2^2+ \eta\left(\sqrt{m}\norm{\bm{\nu}-\bm{\mu}}_2^2+3CK^2\gamma(\mathbb{D}_k+\mathbb{D}_l)\right)^2\right] \nonumber\\
    &= \frac{1}{4} (1-\eta m)
    (\norm{ \bm{\nu}+\bm{\mu}}_2^2- \norm{ \bm{\nu}-\bm{\mu}}_2^2)+\frac{3}{2}\eta  \sqrt{m}  CK^2\gamma(\mathbb{D}_k+\mathbb{D}_l)\left(\norm{ \bm{\nu}+\bm{\mu}}_2+\norm{ \bm{\nu}-\bm{\mu}}_2\right) \nonumber\\
    &= (1-\eta m) \ip{ \bm{\nu}}{\bm{\mu}} +\frac{3}{2}\eta  \sqrt{m}  CK^2\gamma(\mathbb{D}_k+\mathbb{D}_l)\left(\norm{ \bm{\nu}+\bm{\mu}}_2+\norm{ \bm{\nu}-\bm{\mu}}_2\right) \nonumber\\
    &\le 1-\eta  m +6\eta  \sqrt{m}  CK^2\gamma(\mathbb{D}_k+\mathbb{D}_l) \nonumber\\
    &= 1-\eta m \cdot \frac{\sqrt{m}-6CK^2\gamma(\mathbb{D}_k+\mathbb{D}_l)}{\sqrt{m}},
\end{align}
where the last inequality uses $\norm{ \bm{\nu}+\bm{\mu}}_2 \leq 2$ and $\norm{\bm{\nu}-\bm{\mu}}_2\leq 2$.

\noindent \textbf{Case II:} If $\sqrt{m} \norm{ \bm{\nu}+\bm{\mu}}_2 -3CK^2\gamma(\mathbb{D}_k+\mathbb{D}_l)< 0$, we have
\begin{align}
    &~~~\ip{(\vI - \eta  \vA^T \vA) \bm{\nu}}{\bm{\mu}}=\ip{ \bm{\nu}}{\bm{\mu}}-\eta  \ip{\vA \bm{\nu}}{\vA\bm{\mu}} \nonumber\\
    &=\frac{1}{4}\left[\norm{ \bm{\nu}+\bm{\mu}}_2^2 -\eta  \norm{\vA( \bm{\nu}+\bm{\mu})}_2^2-\norm{ \bm{\nu}-\bm{\mu}}_2^2+\eta  \norm{\vA( \bm{\nu}-\bm{\mu})}_2^2\right) \nonumber\\
    &\le \frac{1}{4}\left[\norm{ \bm{\nu}+\bm{\mu}}_2^2 -\norm{ \bm{\nu}-\bm{\mu}}_2^2+\eta \left(\sqrt{m}\norm{ \bm{\nu}-\bm{\mu}}_2 +3CK^2\gamma(\mathbb{D}_k+\mathbb{D}_l)\right)^2\right] \nonumber\\
    &= \frac{1}{4} \norm{ \bm{\nu}+\bm{\mu}}_2^2 -
    \frac{1}{4} (1-\eta m)
    \norm{ \bm{\nu}-\bm{\mu}}_2^2+\frac{1}{4}\eta  3CK^2\gamma(\mathbb{D}_k+\mathbb{D}_l)(3CK^2\gamma(\mathbb{D}_k+\mathbb{D}_l)+2\sqrt{m} \norm{ \bm{\nu}-\bm{\mu}}_2) \nonumber\\
    &\le \frac{1}{4} \norm{ \bm{\nu}+\bm{\mu}}_2^2 +\frac{3}{4}\eta   CK^2\gamma(\mathbb{D}_k+\mathbb{D}_l)(3CK^2\gamma(\mathbb{D}_k+\mathbb{D}_l)+2\sqrt{m} \norm{ \bm{\nu}-\bm{\mu}}_2) \nonumber\\
    &\le \frac{3CK^2\gamma(\mathbb{D}_k+\mathbb{D}_l)}{\sqrt{m}}\left[ \frac{1}{4}\frac{3CK^2\gamma(\mathbb{D}_k+\mathbb{D}_l)}{\sqrt{m}}+\frac{1}{4}\eta  m  \left(\frac{3CK^2\gamma(\mathbb{D}_k+\mathbb{D}_l)}{\sqrt{m}}+2\norm{ \bm{\nu}-\bm{\mu}}_2\right)
    \right] \nonumber\\
    &\le \frac{3CK^2\gamma(\mathbb{D}_k+\mathbb{D}_l)}{\sqrt{m}}\left(\frac{3CK^2\gamma(\mathbb{D}_k+\mathbb{D}_l)}{2\sqrt{m}}+1\right),
\end{align}
where the second inequality we uses $\eta  m \le 1$, and the last inequality uses $\eta  m \le 1$, $\norm{ \bm{\nu}-\bm{\mu}}_2 \le 2$ and $8CK^2 \gamma(\mathbb{D}_k)\le \sqrt{m}$.

Therefore, we obtain
\begin{align}
    \norm{[(\vI - \eta  \vA^T \vA) \vv]_\mathbb{L}}_2
    &=\norm{\vv}_2\ip{(\vI - \eta \vA^T \vA)  \bm{\nu} }{\bm{\mu}}\le 
     \max\left\{s_1,s_2 \right\} \cdot \norm{\vv}_2, \label{eq: differenceterm}
\end{align}
where 
\begin{align}
  s_1=1-\eta m \cdot \frac{\sqrt{m}-6CK^2\gamma(\mathbb{D}_k+\mathbb{D}_l)}{\sqrt{m}},~s_2=\frac{3CK^2\gamma(\mathbb{D}_k+\mathbb{D}_l)}{\sqrt{m}}\left(\frac{3CK^2\gamma(\mathbb{D}_k+\mathbb{D}_l)}{2\sqrt{m}}+1\right).  
\end{align}

\noindent \textbf{Proof of part b)}:
Let 
\begin{equation}
    \vz=\frac{[\vA^T \vx]_\mathbb{L} }{\norm{[\vA^T \vx]_\mathbb{L}}_2} \in \mathbb{D}_l,
\end{equation}
then we have 
\begin{align} 
    \norm{[\vA^T \vx]_\mathbb{L}}_2&=\ip{[\vA^T \vx]_\mathbb{L} }{\vz} =\ip{\vA^T \vx }{\vz}=\ip{\vx}{\vA \vz} \nonumber\\
    &\le \norm{\vx}_2 \norm{\vA \vz}_2\le \norm{\vx}_2(\sqrt{m} +2CK^2\gamma(\mathbb{D}_l)). \label{eq: norsyterm}
\end{align}
with probability at least $1-2\exp(-\gamma^2(\mathbb{D}_l))$, where the first inequality uses Holder's inequality and the last inequality uses Lemma \ref{lm:MatrixDeviationInequality}.

\subsection{Proof of Lemma \ref{lm:Dk}}
\textbf{Proof of part a)}: For any $\vx \in \mathbb{D}_{2k}$, there exist two vectors $\vy,\vz \in \mathbb{D}_{k}$ such that $\vx=(\vy+\vz)/\norm{\vy+\vz}_2$ and $y_i z_i=0, i=1,\ldots,n$, which means the locations of non-zeros for $\vy$ and $\vz$ are different. Then we have
\begin{align}
    \gamma(\mathbb{D}_{2k})&= \textbf{E}\sup_{\vx \in \mathbb{D}_{2k}} |\ip{\vg}{\vx}| \nonumber\\
    &=\textbf{E}\sup_{(\vy+\vz)/\norm{\vy+\vz}_2 \in \mathbb{D}_{2k}} \left|\ip{\vg}{\frac{\vy+\vz}{\norm{\vy+\vz}_2}}\right| \nonumber\\
    &=\textbf{E}\sup_{\vy \in \mathbb{D}_{k}, \vz \in \mathbb{D}_{k}} \left|\ip{\vg}{\frac{\vy+\vz}{\norm{\vy+\vz}_2}}\right| \nonumber\\
    &\le \textbf{E}\sup_{\vy \in \mathbb{D}_{k}} \left|\ip{\vg}{\frac{\vy}{\norm{\vy+\vz}_2}}\right| +\textbf{E}\sup_{\vz \in \mathbb{D}_{k}} \left|\ip{\vg}{\frac{\vz}{\norm{\vy+\vz}_2}}\right| \nonumber\\
    &\le \textbf{E}\sup_{\vy \in \mathbb{D}_{k}} \left|\ip{\vg}{\frac{\vy}{\norm{\vy}_2}}\right| +\textbf{E}\sup_{\vz \in \mathbb{D}_{k}} \left|\ip{\vg}{\frac{\vz}{\norm{\vy}_2}}\right| \nonumber\\
     &= \textbf{E}\sup_{\vy \in \mathbb{D}_{k}} \left|\ip{\vg}{\vy}\right| +\textbf{E}\sup_{\vz \in \mathbb{D}_{k}} \left|\ip{\vg}{\vz}\right| \nonumber\\
     &= 2 \gamma(\mathbb{D}_{k}),
\end{align}
where the first inequality uses the triangle inequality and the second inequality applies the fact that the locations of non-zeros for $\vy$ and $\vz$ are totally different.

\noindent\textbf{Proof of part b)}: For any $\vx \in \mathbb{D}_{k}+\mathbb{D}_{k}$, there exists $\vy,\vz \in \mathbb{D}_{k}$ such that $\vx=\vy+\vz$. So we have
\begin{align}
    \gamma(\mathbb{D}_{k}+\mathbb{D}_{k})&= \textbf{E}\sup_{\vx \in \mathbb{D}_{k}+\mathbb{D}_{k}} |\ip{\vg}{\vx}| \nonumber\\
    &=\textbf{E}\sup_{\vy \in \mathbb{D}_{k}, \vz \in \mathbb{D}_{k}} \left|\ip{\vg}{{\vy+\vz}}\right| \nonumber\\
    &\le \textbf{E}\sup_{\vy \in \mathbb{D}_{k}} \left|\ip{\vg}{{\vy}}\right| +\textbf{E}\sup_{\vz \in \mathbb{D}_{k}} \left|\ip{\vg}{\vz}\right| \nonumber\\
     &= 2 \gamma(\mathbb{D}_{k}).
\end{align}

\noindent \textbf{Proof of part c)}: By the definition of $\mathbb{D}_{k}$, it is clear that   $\mathbb{D}_{k} \subseteq \mathbb{D}_{l}$ when $k\le l$. Therefore, we have $\gamma(\mathbb{D}_{k}) \le \gamma(\mathbb{D}_{l})$.

\end{document}